\documentclass[11pt]{article}
\newcommand{\beq}{\begin{eqnarray}}
\newcommand{\eeq}{\end{eqnarray}}

\newcommand{\be}{\begin{equation}}
\newcommand{\ee}{\end{equation}}
\newcommand{\lwrsim}{\raise0.3ex\hbox{$<$\kern-0.75em\raise-1.1ex\hbox{$\sim$}}}

\def\eq#1{Eq.~(\ref{#1})}

\def\prd#1#2#3{Phys.\ Rev.\ {\bf D#1} (#2) #3}

\def\plb#1#2#3{Phys.\ Lett.\ {\bf B#1} (#2) #3}

\usepackage{axodraw}
\usepackage{amsfonts}
\usepackage{amsmath}
\usepackage{amssymb}
\usepackage{latexsym}
\usepackage{color}
\usepackage{colordvi}
\usepackage{epsfig}
\usepackage{array}
\usepackage{pifont}
\usepackage{geometry}
\newcommand{\ghostSD}{\begin{picture}(150,25)(0,0)
\SetWidth{1.2}
\DashArrowLine(12.5,0)(37.5,0){5}
\DashArrowLine(37.5,0)(75,0){5}
\DashLine(75,0)(112.5,0){5}
\DashArrowLine(112.5,0)(137.5,0){5}
\SetWidth{1}
\Vertex(112.5,0){2}
\GlueArc(75,0)(37.5,0,90){-4}{6}
\GlueArc(75,0)(37.5,90,180){-4}{6}
\CCirc(75,0){5}{Black}{Yellow}
\CCirc(75,37.5){5}{Black}{Yellow}
\CCirc(37.5,0){5}{Black}{Yellow}
\Text(20,-10)[l]{a,k}
\Text(50,15)[l]{d,$\nu$}
\Text(100,-10)[l]{e}
\Text(100,15)[r]{f,$\mu$}
\Text(50,-10)[l]{c,q}
\Text(120,-10)[l]{b,k}
\Text(75,48)[c]{q-k}
\end{picture}}
\newcommand{\ghostDr}{\begin{picture}(100,25)(0,0)
\SetWidth{1.2}
\DashArrowLine(12.5,0)(50,0){5}
\DashArrowLine(50,0)(87.5,0){5}
\CCirc(50,0){5}{Black}{Yellow}
\Text(12.5,-10)[l]{a}
\Text(87.5,-10)[r]{b}
\Text(50,-10)[c]{k}
\end{picture}}
\newcommand{\ghostBr}{\begin{picture}(100,25)(0,0)
\SetWidth{1.2}
\DashArrowLine(12.5,0)(87.5,0){5}
\Text(12.5,-10)[l]{a}
\Text(87.5,-10)[r]{b}
\Text(50,-10)[c]{k}
\end{picture}}
\begin{document}

\title{Is the QCD ghost dressing function finite at zero momentum ?}

\author{ Ph.~Boucaud$^a$, Th. Br\"untjen$^a$, J.P.~Leroy$^a$, A.~Le~Yaouanc$^a$,
\\ A.Y.~Lokhov$^b$,
J. Micheli$^a$, O. P\`ene$^a$ and J.~Rodr\'iguez-Quintero$^c$ }

\date{}

\maketitle
\begin{center}
$^a$Laboratoire de Physique Th\'eorique et Hautes
Energies\footnote{Unit\'e Mixte de Recherche 8627 du Centre National de 
la Recherche Scientifique}\\
{Universit\'e de Paris XI, B\^atiment 211, 91405 Orsay Cedex,
France}\\
$^b$ Centre de Physique Th\'eorique\footnote{
Unit\'e Mixte de Recherche 7644 du Centre National de 
la Recherche Scientifique\\ 
}de l'Ecole Polytechnique\\
F91128 Palaiseau cedex, France\\ 
$^c$ Dpto. F\'isica Aplicada, Fac. Ciencias Experimentales,\\
Universidad de Huelva, 21071 Huelva, Spain.
\end{center}
%\noindent P.A.C.S.: 12.38.Aw; 12.38.Gc; 12.38.Cy; 11.15.H

%\pacs{05.45.Yv,12.38.Aw,11.15.Ha}

%\maketitle

\begin{abstract}
\bigskip

We show that a finite non-vanishing ghost dressing function at zero
momentum satisfies the scaling properties of the ghost propagator Schwinger-Dyson
equation. This kind of Schwinger-Dyson solutions may well agree with  lattice
data and provides an interesting alternative to the widely spread
claim that the gluon dressing function behaves like the inverse
squared ghost dressing function, a claim which is at odds with lattice
data. We demonstrate that, if the ghost dressing function is less  
singular than any power of $p$,  it must be finite non-vanishing
at zero momentum: any logarithmic behaviour is for instance excluded.  
We add some remarks about coupled Schwinger-Dyson analyses.

\end{abstract}

%maketitle

\vspace*{0.5cm}

\begin{flushright}
LPT-ORSAY/06-19 \\
CPHT-RR018.0406 \\
UHU-FT/06-10
\end{flushright}

\section{Introduction}

The infrared behaviour of Landau gauge lattice gluon and ghost
propagators  is an interesting and hot subject. Two main methods
are used: lattice QCD  (LQCD) and  Schwinger-Dyson equations (SDE) in which we
include related methods as RGE, etc. In
ref.~\cite{Boucaud:2005ce} we  have shown that a combination of both
methods  is extremely enlightening  as it combines the advantages of 
lattice QCD's full control of errors and SDE's analytical character.

We only consider the particularly simple ghost propagator SDE :
 \vspace{\baselineskip}
\begin{small}
\beq
\left(\ghostDr\right)^{-1}% 
=
\left(\ghostBr\right)^{-1}%
- 
\ghostSD % 
\nonumber
\eeq\end{small}% 

We have studied the discrepancy between  LQCD data and  a widely
 spread belief: the ghost propagator SDE is claimed to imply a gluon
 dressing function behaving like the inverse squared ghost one. 
 In ref.~\cite{Boucaud:2005ce} we have reconsidered the scaling 
 properties of the SDE and found three possible ways out of this problem
 (which are summarised in table 1 of that paper). The 
 first one is to assume a singular behaviour  of the 
 ghost-gluon vertex in the deep infrared. A second possibility implies a
 very singular ghost dressing function which is excluded by LQCD. 
 The third one is to
 assume  that the ghost dressing function is less singular in the
 infrared that any power of $p$. In view of the general belief that
 the ghost dressing function was strongly  singular we had not paid 
 in ref.~\cite{Boucaud:2005ce} attention to  
 the third one. 
  
  Very recently,  Sternbeck et al. \cite{Sternbeck:2005re} have
  produced two new evidences: i) the ghost-gluon vertex seems
  not to be singular, ii) the ghost dressing function seems to behave
  at most like $\log p$ in the infrared. These two evidences, taken together,
  strongly encourage us to consider now seriously  the third above-mentioned
  solution.  This is the aim of the present letter. 

  To our surprise we found that one can {\it  demonstrate} from
   the scaling analysis of the ghost propagator SDE  the impossibility of 
   a  $\log p$
   behaviour or any other behaviour which is less divergent than
   any power of $p$: under these conditions, the ghost dressing function  
   necessarily has a {\it finite non-vanishing} limit at zero momentum. This is at odds
   with a very general belief that the ghost dressing function is
   divergent. The proof will be displayed in section~\ref{SDC}. We will shortly 
discuss published results about coupled gluon and ghost SDE in 
   section \ref{gluon}.
   
   \section{Notations and summary of up-to-date lattice results}
   \label{notations}
   
   We use the following notations~\cite{Boucaud:2005ce}:

\beq\label{Def} \widetilde{\Gamma}_{\mu}(-q,k;q-k)&=&q_\mu
H_1(q,k)+(q-k)_{\mu}H_2(q,k) \nonumber \\ \left(
F^{(2)}\right)^{ab}(k^2) &=& - \delta^{ab} \ \frac{F(k^2)}{k^2} 
\nonumber \\ \left( G^{(2)}_{\mu\nu}\right)^{ab}(k^2) &=& 
\delta^{ab} \ \frac{G(k^2)}{k^2}  \left( \delta_{\mu\nu}-\frac{k_\mu
k_\nu}{k^2}\right) , \eeq
where $G^{(2)}$ and $F^{(2)}$ are respectively the gluon and ghost
propagators, $G$ and $F$ are respectively the gluon and ghost dressing
functions and where $\widetilde{\Gamma}_{\mu}(-q,k;q-k)$ is the
ghost-gluon vertex ( $k$ and $q$ are the momenta of the incoming and
outgoing ghosts and $q-k$ the momentum of the gluon) .

 Following for simplicity the common, convenient, but not really
  justified, assumption of a power-law behaviour of the propagators in
  the deep infrared, we define 
 \beq\label{FG}
 F(k^2) &\sim& \left( \frac {k^2}{\nu}\right)^{\alpha_F} ,\
G(k^2) \sim \left( \frac {k^2}{\lambda}\right)^{\alpha_G}.
\eeq
In ref.~\cite{Boucaud:2005ce} we have also defined an infrared exponent 
$\alpha_\Gamma$ for the vertex funtion $H_1$ ($\alpha_\Gamma <0$ 
means a singular infrared behaviour).

Using the ghost propagator SDE equation it is often claimed that $2\alpha_F + \alpha_G=0$.
This belief is so strong that one often uses only one parameter $\kappa = -\alpha_F
=\alpha_G/2$. However, as we will see in more details, everybody agrees that 
$\alpha_G$ is close~\footnote{For example, in many SDE approaches it is 
found~\cite{Alkofer:2000wg} $\alpha_G \simeq 1.18$} to 1 and it becomes now 
clear~\footnote{ One may wonder why many power law  fits have given  negative 
$\alpha_F$. Our own fit in  ref.\cite{Boucaud:2005ce} (table 2) has produced negative values 
very close to zero, but the errors were clearly underestimated. 
Presumably the systematic one, due to the functional form chosen for the 
fit, has not been properly taken into account.  This is also the case in 
several other published results.} 
that $\alpha_F $ is close to zero. 
Then {\it the relation $2\alpha_F + \alpha_G=0$ is not satisfied~\cite{
Boucaud:2005ce, Sternbeck:2005re}  and the arguments which support it
have to be reconsidered}. 

\paragraph{The lattice gluon propagator.} Several SDE studies~(\cite{Alkofer:2000wg} and references therein)
predict a vanishing  zero momentum propagator while, as discussed in
ref.~\cite{Boucaud:2006pc},  a gluon propagator converging continuously
to a non-zero value at vanishing momentum is a rather general
lattice result (in particular, the authors  of
ref.~\cite{Bonnet:2001uh} obtain a non-vanishing infrared limit 
for the gluon propagator at a lattice volume of around 2000 fm$^4$).
Therefore our preferred solution~\footnote{Let us recall however that there is still 
a problem coming from the Slavnov-Taylor identity for the three gluon-vertex: we have shown 
that, the vertices being regular when one momentum tends to zero, it implies $\alpha_G < 1$. Is it possible to avoid 
any contradiction by assuming, as done by Cornwall~\cite{Cornwall}, a non-regular behaviour for 
the longitudinal part of the three-gluon vertex? This deserves more investigation.} is 
\beq
\alpha_G=1.
\eeq

But, even if we relax this relation and assume a
vanishing gluon propagator with $\alpha_G >1$, the
solution with a finite ghost dressing function at zero momentum 
remains possible as we shall see. 

%\label{latres}
\paragraph{The lattice ghost propagator.} Very recent lattice estimates~\cite{Sternbeck:2005re} seem to  point
towards a ghost dressing function  rather close to the perturbative
behaviour: the dressing  function only shows, if any, a logarithmic
dependence on the 
momentum (see fig. 2 of ref.~\cite{Sternbeck:2005re}).

\begin{figure}
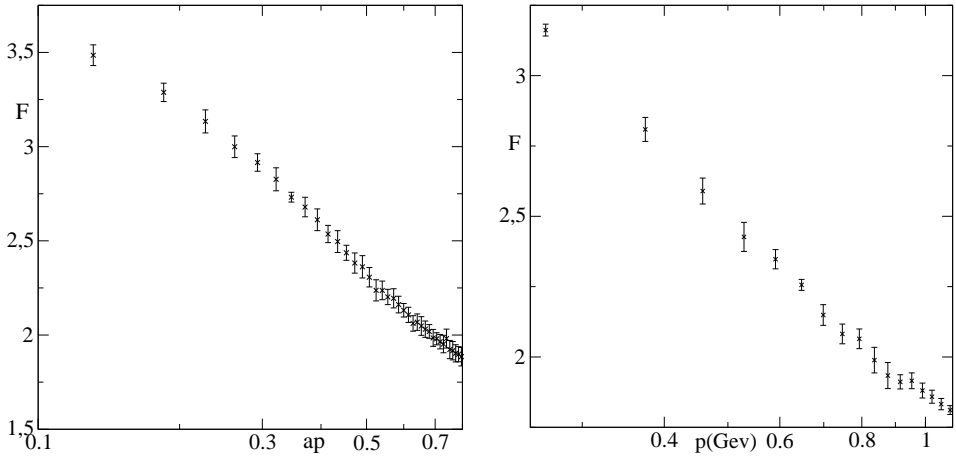

\begin{center}
\begin{tabular}{cc}
\mbox{\epsfig{file=SU2.eps,height=6cm}} 
&
\mbox{\epsfig{file=SU3.eps,height=6cm}}
\end{tabular}
\end{center}
\caption{$F(p)$ from lattice simulations for $SU(2)$ (left, $Vol=48^4$, $\beta = 2.3$) and
$SU(3)$ (right,$Vol=32^4$, $\beta = 5.8$). $\beta=2.3$ for $SU(2)$ has been chosen to 
guarantee that the string tension in lattice units is close to that of $\beta=5.8$ for $SU(3)$.}
\label{Fig4}
\end{figure}

We confirm these results.
In Fig.~\ref{Fig4} the ghost dressing function is plotted as a
function of $\log(p)$ for small values of the momenta. These plots were
obtained from lattice simulations at $\beta=5.8$ and a volume $32^4$
in the $SU(3)$ case and at $\beta=2.3$ and a volume $48^4$ in the SU(2)
case. It is clear from these plots that $F(p)$ does not exhibit any
power law: $\alpha_F=0$. For SU(3) $F(p)$ is approximately linear in
$\log(p)$ and for $SU(2)$ it has even a smoother behavior (In this case one
obtains a good fit of the data with a function  $C (\log\vert p \vert)
^\gamma$ and $\gamma \thickapprox 0.4$). 

\paragraph{The ghost-gluon vertex.} In ref.~\cite{Sternbeck:2005re}, the authors did not find any evidence for a
singularity in the case of a vanishing gluon momentum. Let us remark
that this particular kinematical configuration isolates the form factor $H_1$
(see Eq.(\ref{Def})) which enters in the SD equation (It is worth 
recalling that in perturbation theory $H_1(q,0)+H_2(q,0)$ is equal to 1 although
$H_1(q,0)$ is not~\cite{Chetyrkin:2000dq}). Our dimensional 
analysis of the ghost SDE (see next section \ref{SDC}) invokes a different 
kinematical configuration for this form factor, $H_1(q,k \to 0)$ 
instead of $H_1(k,k)$. The non-singular behaviour they found as $k$ tends to 0 
excludes however the singularity of the ghost-gluon vertex that we 
proposed in ref.~\cite{Boucaud:2005ce} 
as our favoured solution to reconcile the ghost propagator SDE and the lattice  inspired  relation
$2\alpha_F + \alpha_G > 0$.

\paragraph{In  conclusion} lattice simulations  show a strong evidence
that $\alpha_G = 1$ and $2\alpha_F + \alpha_G >0$, far from zero.
Now we have a fair indication that $\alpha_F = 0$ and about the regularity of the 
vertex form factor involved in the ghost SDE. This leads us to revisit, in next section, 
the case in column 4 of table 1 in ref.~\cite{Boucaud:2005ce} for ghost and gluon 
propagators and vertices satisfying the scaling properties of ghost SDE.

\section{Ghost SDE: the case $\alpha_F = 0$}\label{SDC}

We will now demonstrate that $F(0)$ is finite non-vanishing for $\alpha_F=0$. We 
will exploit the constraints, summarized in table 1 of ref.~\cite{Boucaud:2005ce},
 between the infrared exponents $\alpha_F$, $\alpha_G$, $\alpha_\Gamma$, from 
 the ghost SDE. The IR convergence of the loop integral in the ghost SDE implies the 
 two conditions:
\beq\label{cond_IR}
 \alpha_F+\alpha_\Gamma > -2  \      ,     \  \alpha_G+\alpha_\Gamma > -1 \ .
\eeq
Then the dimensional consistency of ghost SDE at small momenta leads to only three allowed 
cases:

i) $\alpha_F\neq 0$ and $\alpha_F + \alpha_G  +\alpha_\Gamma <1 \Longrightarrow 2\alpha_F+
\alpha_G+\alpha_\Gamma=0$

ii) $\alpha_F\neq 0$ and $\alpha_F + \alpha_G  +\alpha_\Gamma \geq 1 
\Longrightarrow \alpha_F = -1$

iii) $\alpha_F = 0$ and $\alpha_G  +\alpha_\Gamma \geq 1$ \ does not require any further constraint.

We shall look, in the following, at the consequences of the third case. It includes in particular
$\alpha_G = 1$ and $\alpha_\Gamma = 0$ which is 
favoured by lattice simulations (see section (\ref{notations})~). Nevertheless we shall not
suppose, in the following derivation, anything more than $\alpha_G + \alpha_\Gamma \geq 1$
and conditions (\ref{cond_IR}). This leaves open, for example, the possibility that the
gluon propagator goes to zero in the IR limit, the vertex remaining finite or singular.

Of course, even with $\alpha_F = 0$, we cannot exclude {\it a priori} the possibility that $F(k)$ 
diverges or tends to zero more slowly than any power of $k$ when $k\rightarrow 0$. We shall 
however prove  that this is not allowed: $F(k)$ remains finite in this limit provided that the 
two following conditions are satisfied:

\beq\label{cond}
\alpha_F = 0  \      ,      \   \alpha_G+ \alpha_\Gamma \geq 1
\eeq

Writing the subtracted bare SD equation for two scales $\lambda k$ and $\kappa \lambda k$
(see Eq.(11) of ref.~\cite{Boucaud:2005ce}) one obtains:

\beq\label{SDeq}
\frac{1}{F(\lambda k)}-\frac{1}{F(\kappa \lambda k)}=g_{B}^2 N_c\int \frac{d^4 q}{(2\pi)^4}
\Big(\frac{F(q^2)}{q^2} \left(\frac{(k\cdot q)^2}{k^2}-q^2\right) \nonumber\\
\times\Big[\frac{G((q-\lambda k)^2)H1(q,\lambda k)}{(q-\lambda k)^2}-
\frac{G((q-\kappa \lambda k)^2)H1(q,\kappa \lambda k)}{(q-\kappa \lambda k)^2}
\Big]\Big) \ ,
\eeq
where $\lambda$ is a parameter which we shall use to study the IR ($\lambda \to 0$)
dimensional behaviour of $F$; $\kappa$ is a fixed number, $0< \kappa < 1$,
needed to write a subtracted equation ensuring the UV convergence. 
%The integral in the r.h.s. is IR convergent if conditions(\ref{cond_IR}) are satisfied. 
It was shown in ref.~\cite{Boucaud:2005ce} that the r.h.s. of \eq{SDeq} is the sum of 
two terms behaving respectively as
$\lambda^{2{\rm Min}(\alpha_F+\alpha_G+\alpha_\Gamma,1)}$ and $\lambda^2$ when 
$\lambda\rightarrow 0$.
So it behaves as $\lambda^2$ when the conditions (\ref{cond}) are satisfied.
For any $\kappa$ there is a value of $\lambda$ and $c$ such that 
$\forall \lambda' \leq \lambda$ we have
$\vert \frac{1}{F(\lambda' k)}-\frac{1}{F(\kappa \lambda' k)}\vert 
\leq c \lambda'^2$, thus:

\beq
\vert \frac{1}{F(\lambda k)}-\frac{1}{F(\kappa \lambda k)}\vert &\leq& c \lambda^2 \nonumber \\
&\vdots& \nonumber \\ 
\vert \frac{1}{F(\kappa^{n-1}\lambda k)}-\frac{1}{F(\kappa^n \lambda k)}\vert
&\leq& c \lambda^2 \kappa^{2(n-1)} 
\eeq
which implies:
\beq\label{bound}
\vert \frac{1}{F(\lambda k)}-\frac{1}{F(\kappa^n \lambda k)}\vert \leq c
 \frac{1-\kappa^{2n}}{1-\kappa^2}\lambda^2.
\eeq
So $F\rightarrow\infty$ when $\lambda\rightarrow 0$ is excluded because taking the limit of the above
expression when $n\rightarrow\infty$ we should have $\vert \frac{1}{F(\lambda k)}\vert
\leq c \frac{1}{1-\kappa^2}\lambda^2$ and F would diverge as or more rapidly than 
$\frac{1}{\lambda^2}$ implying $\alpha_F \leq -1$ in contradiction with the hypothesis 
$\alpha_F=0$.
Let us remark that $F\rightarrow 0$ is also excluded: Eq. (\ref{bound}) implies 
$\vert \frac{1}{F(\kappa^n \lambda k)}\vert \leq \vert \frac{1}{F(\lambda k)}
\vert+c\frac{1-\kappa^{2n}}{1-\kappa^2}\lambda^2$ and $\frac{1}{F(\kappa^n \lambda k)}$ 
cannot tend to  infinity when $n\rightarrow\infty$.
This completes the proof.  Notice that we have used bare Green functions and
couplings, everything remains however exactly the same if we replace them by 
renormalized ones.  

\textbf{This is our main result: If $\alpha_F = 0$ the ghost dressing function has to  be finite
and $\neq 0$ in  the IR limit.} This solution is compatible with our knowledge from lattice
simulations about the behavior of the ghost dressing function and ghost-gluon vertex. 
Of course, the current lattice simulations cannot yet exclude a smooth divergence which 
the preceding dimensional analysis forbids. A detailed numerical study of the ghost propagator 
in the deep IR is strongly needed.

\section{Remarks about coupled gluon and ghost SDE solutions}
\label{gluon}

The combination of the scaling analysis of ghost SDE and lattice predictions
appears to be very restrictive concerning the low-momentum behaviour of gluon 
and ghost propagators. Such a behaviour must be a solution of the combined SDE 
for both gluon and ghost propagators. The schemes followed to solve the combined 
SDE's have often led to $2 \alpha_F + \alpha_G = 0$ and $\alpha_G \gtrapprox 1$ 
(hence a strongly divergent ghost dressing function). However, a two-loop analysis~\cite{Bloch:2003yu} 
proves to be much less restrictive in constraining $\alpha_G$. Our findings require to 
reconsider these approaches  by taking into due account the special case 
$\alpha_F=0$. 

In a recent paper~\cite{Aguilar:2004kt} Aguilar and Natale found $\alpha_G=1$ and $\alpha_F=-0.04$,
not far from our present conclusions and deserving a closer comparison.
They followed the Cornwall~\cite{Cornwall}
prescription for the trilinear gluon vertex and solved the coupled equations for the ghost 
and gluon propagators in the Mandelstam approximation~\cite{Mandelstam}.  Concerning the ghost 
dressing function, in spite of the fact that they find it slowly power-like divergent,
it remains flat till very small momenta. This last point is in contradiction with
the lattice results in ref.~\cite{Sternbeck:2005re} and ours in Fig.~\ref{Fig4}, where 
$F(k)$ is not at all so flat and shows a logarithmic enhancement as the momentum decreases. 
Of course, if power-like divergences are excluded, the arguments presented in section 
\ref{SDC} imply a flat dressing function in a small momentum range presumably not yet 
reached by current lattice analyses. 
  
\begin{figure}[htb]
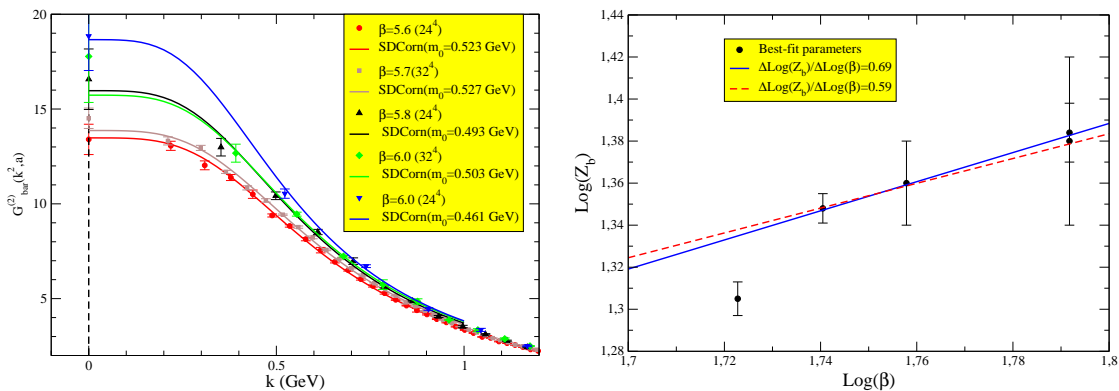

\begin{center}
\begin{tabular}{cc}
\mbox{\epsfig{file=G2-all.eps,height=5.1cm}} &
\mbox{\epsfig{file=logzb.eps,height=5.1cm}}
\end{tabular}
\end{center}
\caption{\small Best fits to the lattice data (left) and 
Log-log plot of $Z_b$ in terms of the lattice bare gauge
coupling parameter $\beta$ (right). The solid blue (dotted red) line shows a fit to a 
linear formula where the slope is free to be fitted (fixed by one-loop perturbation  
theory in \eq{one-loop}~).}
\label{Fig1}
\end{figure}

To compare quantitatively their gluon propagator with LQCD~\cite{Boucaud:2003xi}, we have applied 
the simple parametrization they proposed: 
\beq\label{lattmodel}
G^{(2)}_{\rm bare}(q^2;a,L) \ = \  
\frac {Z_b(a)} {q^2 + \displaystyle \frac{m_0(L)^4}{q^2+m_0(L)^2}} \ + {\cal O}(a,1/L) \ ,
\eeq 
where $a$ stands for the lattice spacing and $L$ for the lattice length. 
In ref.~\cite{Boucaud:2003xi}, the gluon propagator was estimated from $24^4$ lattices at 
$\beta=5.6,5.8,6.0$ and $32^4$ lattices at $\beta=5.7$ and $\beta=6.0$
and analyzed through an instanton liquid model that  failed in describing the
very low momentum range ($q < 0.4$GeV).
In fig.~\ref{Fig1}, we plot the curves corresponding to the best-fit parameters $m_0$ and $Z_b$
collected in table \ref{table1}. The parametrization \eq{lattmodel}
matchs pretty well the lattice data~\footnote{The masses we obtain differ from the one quoted in
ref.~\cite{Aguilar:2004kt} but these depend on a parameter, $\Lambda$, which in their approach can be
varied.}. Moreover one knows 
that, at the leading log,

\beq\label{one-loop}
d(\log(Z_b(a))=\frac{13}{22}d(\log\beta) \ .
\eeq 
Performing a linear fit of $\log(Z_b)$ as a function of
$\log(\beta)$ for $\beta \geq 5.7$ we obtain a slope approximately equal to 0.69 which has to 
be compared to
$\frac{13}{22}=0.59$. That result is unexpectedly good for the large lattice spacings we take 
in consideration.

\begin{table}
\begin{center}
\begin{tabular}{||c|c|c||}
\hline
\hline
lattice & $m_0$ (GeV) & $Z_b(a)$ \\
\hline 
5.6 ($24^4$) & 0.523(2) & 3.69(1)  \\
5.7 ($32^4$) & 0.527(1) & 3.85(1)  \\
5.8 ($24^4$) & 0.493(7) & 3.88(3)  \\
6.0 ($32^4$) & 0.503(4) & 3.99(2)  \\
6.0 ($24^4$) & 0.461(16) & 3.97(6) \\
\hline 
\hline 
\end{tabular}
\end{center}
\caption{\small best-fit parameters.}
\label{table1}
\end{table}

\section{Conclusions}
The main result we presented in this brief note was to emphasize the interest of  a general class
of SDE solutions where $\alpha_F \thickapprox 0$. This solution has the advantage of being compatible with other
convincing lattice results namely: $\alpha_G=1$ and $2\alpha_F+\alpha_G > 0$. It is also
compatible with a still uncertain result concerning the regularity of the ghost-gluon vertex.
We have proven that if $\alpha_F=0$ the ghost dressing function must be finite in the
IR limit.  Of course one would need measures
on larger volumes in order to test the finiteness of the ghost propagator in
the limit $k\rightarrow 0$.

We have discussed some results from published coupled ghost and gluon SDE
solutions and also shown that the lattice gluon propagator data at low 
momenta can be well described by the very simple parametrisation  
\eq{lattmodel} inspired by a recent gluon SDE analysis.


\begin{thebibliography}{9}
{\small
%----------------------------------------
%\cite{Boucaud:2005ce}
\bibitem{Boucaud:2005ce}
  P.~Boucaud {\it et al.},
  %``The infrared behaviour of the pure Yang-Mills Green functions,''
  arXiv:hep-ph/0507104.
  %%CITATION = HEP-PH 0507104;%%
%\cite{Lokhov:2006es}
%\bibitem{Lokhov:2006es}
  A.~Y.~Lokhov {\it et al.},
  %``Non-perturbative approach to the Landau gauge gluodynamics,''
  arXiv:hep-lat/0601007.
  %%CITATION = HEP-LAT 0601007;%%
  %----------------------------------------
%\cite{Sternbeck:2005re}
\bibitem{Sternbeck:2005re}
  A.~Sternbeck, E.~M.~Ilgenfritz, M.~Muller-Preussker and A.~Schiller,
  %``Landau gauge ghost and gluon propagators and the Faddeev-Popov operator
  %spectrum,''
  Nucl.\ Phys.\ Proc.\ Suppl.\  {\bf 153} (2006) 185
  [arXiv:hep-lat/0511053].
  %%CITATION = HEP-LAT 0511053;%%
%----------------------------------------
%\cite{Alkofer:2000wg}
\bibitem{Alkofer:2000wg}
  R.~Alkofer and L.~von Smekal,
  %``The infrared behavior of QCD Green's functions: Confinement, dynamical
  %symmetry breaking, and hadrons as relativistic bound states,''
  Phys.\ Rept.\  {\bf 353} (2001) 281
  [arXiv:hep-ph/0007355].
  %%CITATION = HEP-PH 0007355;%%
%---------------------------------------
%\cite{Boucaud:2006pc}
\bibitem{Boucaud:2006pc}
  P.~Boucaud {\it et al.},
  %``Short comment about the lattice gluon propagator at vanishing momentum,''
  arXiv:hep-lat/0602006.
  %%CITATION = HEP-LAT 0602006;%%
%---------------------------------------
%\cite{Chetyrkin:2000dq}
\bibitem{Chetyrkin:2000dq}
  K.~G.~Chetyrkin and A.~Retey,
  %``Three-loop three-linear vertices and four-loop MOM beta  functions in
  %massless QCD,''
  arXiv:hep-ph/0007088.
  %%CITATION = HEP-PH 0007088;%%
%-----------------------------------------
%\cite{Boucaud:2003xi}
\bibitem{Boucaud:2003xi}
  P.~Boucaud, F.~De Soto, A.~Le Yaouanc, J.~P.~Leroy, J.~Micheli, O.~Pene and J.~Rodriguez-Quintero,
  %``Evidences for instantons effects in Landau lattice Green functions,''
  Phys.\ Rev.\ D {\bf 70} (2004) 114503
  [arXiv:hep-ph/0312332].
  %%CITATION = HEP-PH 0312332;%%
%----------------------------------------
%\cite{OPE}
\bibitem{OPE}
        Ph. Boucaud, A. Le Yaouanc, J.P. Leroy, J. Micheli, 
        O. P\`ene, J. Rodriguez-Quintero, \plb{493}{2000}{315};
%\bibitem{OPEOne} 
        Ph. Boucaud,A. Le Yaouanc, J.P. Leroy, J. Micheli, 
        O. P\`ene, J. Rodriguez-Quintero, \prd{63}{2001}{114003} 
%---------------------------------------
%\cite{Bonnet:2001uh}
\bibitem{Bonnet:2001uh}
  F.~D.~R.~Bonnet, P.~O.~Bowman, D.~B.~Leinweber, A.~G.~Williams and J.~M.~Zanotti,
  %``Infinite volume and continuum limits of the Landau-gauge gluon
  %propagator,''
  Phys.\ Rev.\ D {\bf 64} (2001) 034501
  [arXiv:hep-lat/0101013].
  %%CITATION = HEP-LAT 0101013;%%
%----------------------------------------
%\cite{Bloch:2003yu}
\bibitem{Bloch:2003yu}
  J.~C.~R.~Bloch,
  %``Two-loop improved truncation of the ghost-gluon Dyson-Schwinger  equations:
  %Multiplicatively renormalizable propagators and  nonperturbative running
  %coupling,''
  Few Body Syst.\  {\bf 33}, 111 (2003)
  [arXiv:hep-ph/0303125];
  %%CITATION = HEP-PH 0303125;%%
%\cite{Bloch:2003sk}
%\bibitem{Bloch:2003sk}
  J.~C.~R.~Bloch, A.~Cucchieri, K.~Langfeld and T.~Mendes,
  %``Propagators and running coupling from SU(2) lattice gauge theory,''
  Nucl.\ Phys.\ B {\bf 687} (2004) 76
  [arXiv:hep-lat/0312036].
  %%CITATION = HEP-LAT 0312036;%% 
%----------------------------------------
%\cite{Aguilar:2004kt}
\bibitem{Aguilar:2004kt}
  A.~C.~Aguilar and A.~A.~Natale,
  %``A dynamical gluon mass solution in Mandelstam's approximation,''
  Int.\ J.\ Mod.\ Phys.\ A {\bf 20} (2005) 7613
  [arXiv:hep-ph/0405024];
  %%CITATION = HEP-PH 0405024;
%%%\cite{Aguilar:2004sw}
%\bibitem{Aguilar:2004sw}
% A.~C.~Aguilar and A.~A.~Natale,
  %``A dynamical gluon mass solution in a coupled system of the  Schwinger-Dyson
  %equations,''
  JHEP {\bf 0408} (2004) 057
  [arXiv:hep-ph/0408254].
  %%CITATION = HEP-PH 0408254;%%
%----------------------------------------
%\cite{Cornwall}
\bibitem{Cornwall}
J.M. Cornwall, Phys. Rev. D {\bf 26} (1982) 11453; 
J.M. Cornwall and J. Papavassiliou, Phis. Rev. D {\bf 40} (1989) 3474.
%----------------------------------------
%\cite{Mandelstam}
\bibitem{Mandelstam}
S. Mandelstam, Phys. Rev. D {\bf 20} (1979) 3223.
%-----------------------------------------


}
\end{thebibliography}
\end{document}